# Structural phase separation in La$_{2/3}$Ba$_{1/3}$MnO$_3$: macroscopic and microscopic aspects

A. B. Beznosov, E. L. Fertman*, P. P. Pal-Val, and L. N. Pal-Val

*B. I. Verkin Institute for Low Temperature Physics and Engineering, National Academy of Sciences of Ukraine, 47 Lenin Ave., 61103 Kharkov, Ukraine*

C. Ritter

*Institut Laue-Langevin, Boite Postale 156, 38042 Grenoble Cedex 9, France*

M. Rossell

*EMAT, University of Antwerp, Groenenborgerlaan 171, B-2020 Antwerpen, Belgium*

D. D. Khalyavin

*Institute of Solid State and Semiconductor Physics, National Academy of Sciences of Belarus, 17 P. Brovka street., 220072 Minsk, Belarus*

The structural phase transformation in the La$_{2/3}$Ba$_{1/3}$MnO$_3$ perovskite manganite was studied by ultrasound, neutron diffraction and electron microscopy techniques. Minima of the velocity $v$ and maxima of the decrement $\delta$ of the ultrasound point out the structural transition temperature $T_s$=200 K. The hysteretic temperature behavior and giant alteration of $v$ and $\delta$ present an evidence of martensitic character of the phase transition at a macroscopic level. The martensitic transformation in La$_{2/3}$Ba$_{1/3}$MnO$_3$ is tightly connected with electronic transitions in both coexisting high symmetry $R\bar{3}c$ and low symmetry *Imma* crystal phases. Neutron diffraction data show that the electronic structure of the compound exhibits anomalous behavior in the vicinity of $T_s$: temperature dependences of the Mn-O-Mn bond angles and Mn-O bond lengths have singularities. Parameters of the coexisting crystalline phases studied have different character of temperature correlation with parameters of the magnetic subsystem of the compound.

# I. INTRODUCTION

Interest in mixed valence manganites with general formula $R_{1-x}A_xMnO_3$ ($R$ = rare-earth cation, $A$ = Ca, Ba, Sr) is caused by the colossal magnetoresistance effect (CMR) they exhibit.[1,2,3] The strong interaction between charge carriers, localized spins, and lattice degrees of freedom is a characteristic feature of these materials, which results in a great variety and complexity of physical phenomena and phase transitions observed. Recently attention has focused on phase-separation phenomena which play a key role in the physics of this class of materials.[2, 4, 5, 6, 7] In particular, the coexistence of metallic and insulating phases leads to a percolative metal-insulator transition and unusually high magnetoresistance (CMR).[2, 4, 5, 8]

Phase separation in manganites may have two origins: 1) electronic (microscopic) phase separation between phases with different charge carriers densities, that results in nanometer scale coexisting clusters, and 2) disorder-induced phase separation with percolative characteristics between equal-density phases, driven by disorder near first-order phase transitions.[2, 9, 10, 11, 12, 13, 14] The later leads to coexisting clusters as large as a micrometer in size. It has become clear by now that first-order phase transformations, which result in the coexistence of two crystalline phases in a wide temperature range (so-called martensitic transformations), play an important role in the physics of manganites. Strong electron correlations and interaction between the coexisting crystalline phases form unique properties of a number of manganites.[9, 11, 12, 15] Physical properties of martensitic systems are governed by the long-range deformations of the crystal lattice, the accommodation strains, which are induced as a result of the nucleation of the martensitic particles within the parent high temperature phase.[16]

Considerable attention is given to a martensitic character of the charge ordering (CO) transition as the transport properties of this class of manganites are governed in many respects by the charge-disordered insulating phase stabilized at low temperature by virtue of martensitic accommodation strain.[8] The martensitic nature of a magnetic-field-induced CO insulating antiferromagnetic to ferromagnetic phase transition was reported to lead to unusual relaxation phenomena in these compounds, such as magnetic-field-induced ultrasharp magnetization steps.[11, 12]

The first order structural transformations are still little studied, although experiments on $La_{1-x}Sr_xMnO_3$ [17, 18, 19, 20] and $La_{1-x}Ba_xMnO_3$ (x=0.2, 0.33)[14, 21, 22, 23] revealed many interesting peculiarities of their behavior, such as the strong dependence of the structural transition temperature $T_s$ on magnetic field and hydrostatic pressure and the wide temperature range of the phases coexistence. The crystal structure of $La_{1-x}Sr_xMnO_3$ manganites has been studied extensively, data about structural features in other manganites are still quite scarce.

Recently it was revealed that the $La_{2/3}Ba_{1/3}MnO_3$ compound experienced a structural phase transition at $T_S = 200$ K.[14, 23] This transition possesses certain features of a martensitic transformation and causes a crystal phase separation in the compound: below the room temperature the substance represented a mixture of the high temperature rhombohedral crystal phase $R\bar{3}c$ (austenite) and the low temperature orthorhombic phase *Imma* (martensite).[14, 23]

In the present paper we have performed the ultrasound, neutron diffraction and electron microscopy study of the $La_{2/3}Ba_{1/3}MnO_3$ manganite. The particular goal was to reveal a correlation between the temperature behavior of the electronic structure and crystal lattice characteristics in the region of martensitic transformation.

## II. EXPERIMENTAL DETAILS

Samples for the experiments were made from polycrystalline pellets of the $La_{2/3}Ba_{1/3}MnO_3$ compound which were prepared using standard solid-state reaction with stoichiometric amounts of powders $La_2O_3$, $BaCO_3$, and $Mn_2O_3$; the details were published earlier.[14] The sample quality was confirmed by x-ray and neutron diffraction study.

The temperature dependences of the ultrasound velocity *v(T)* and decrement *d(T)*, were obtained by the two-component oscillator technique for longitudinal standing waves[24] at the frequencies around 70 kHz between 5 and 340 K.

Neutron powder-diffraction data were obtained using the D1B and D2B diffractometers of the Institute Laue-Langevin (Grenoble) equipped with an ILL cryofurnace. The experiments were performed at wavelengths λ=2.52 Å and λ=1.594 Å, respectively, and the data of the latter were refined using the program FULLPROF.[25] Details of the experiment and spectra refinement are published in Ref. [14].

Electron microscopy study was performed at the University of Antwerp, including room temperature and low temperature experiments. Electron diffraction (ED) patterns were obtained using the Philips CM20 microscope at 300 K and at 100 K. A liquid nitrogen cooling holder was used to decrease the temperature of the samples to 100 K. High-resolution electron microscopy (HREM) studies were performed using a JEOL 4000EX microscope operated at 400 kV.

## III. ELASTIC CHARACTERIZATION

The measurement of the elasticity can give direct information on structural phase transitions in solids. However, elastic properties of CMR compounds remain weakly investigated, although the role of the lattice in determining their properties is generally recognized. Publications on the sound propagation in manganites are not numerous, for example Refs. [26, 27, 28], and only some of them contain information on sound damping. To our knowledge there are no papers dealing with the temperature evolution of the elasticity of $La_{2/3}Ba_{1/3}MnO_3$.

The study of the elastic properties of the $La_{2/3}Ba_{1/3}MnO_3$ compound performed in the present work brought out a number of features (Fig. 1), which are characteristic for martensitic transformations.[29] With decreasing temperature from 340 K to 190 K the ultrasound velocity $v$ strongly decreases from 3.433 km/s to 2.746 km/s. A deep minimum of $v(T)$ is observed at 190 K on cooling and at 202 K on heating (Fig.1, upper panel). The temperature dependences of the decrement $\delta(T)$ on cooling and heating show the corresponding maxima (Fig.1, lower panel). Further cooling below 190 K results in a giant growth of sound velocity $v$ up to 3.969 km/s. At the same time the decrement $\delta$ decreases from 0.02 to 0.005.

These clear anomalies of the temperature dependences of elastic characteristics are caused by the structural phase transition from the $R\bar{3}c$ to the *Imma* space symmetry lattice. The temperature hysteresis of ultrasound properties reflects the 1$^{st}$ order character of the transition.

We suppose, that the specific anomalies of the elastic properties mentioned above give a macroscopic evidence of the martensitic character of the structural phase transition observed. The latter results in the coexistence of two phases with different crystal structures, which are stabilized by the induced inner strain. The coexisting phases form a microstructure of the substance, which can probably represent a sort of the "red-cabbage structure".[30] Marked dips on the temperature dependences of the sound velocities on heating and cooling (Fig.1, upper panel) define a difference of the transition temperatures for the direct and reverse martensitic transformations.

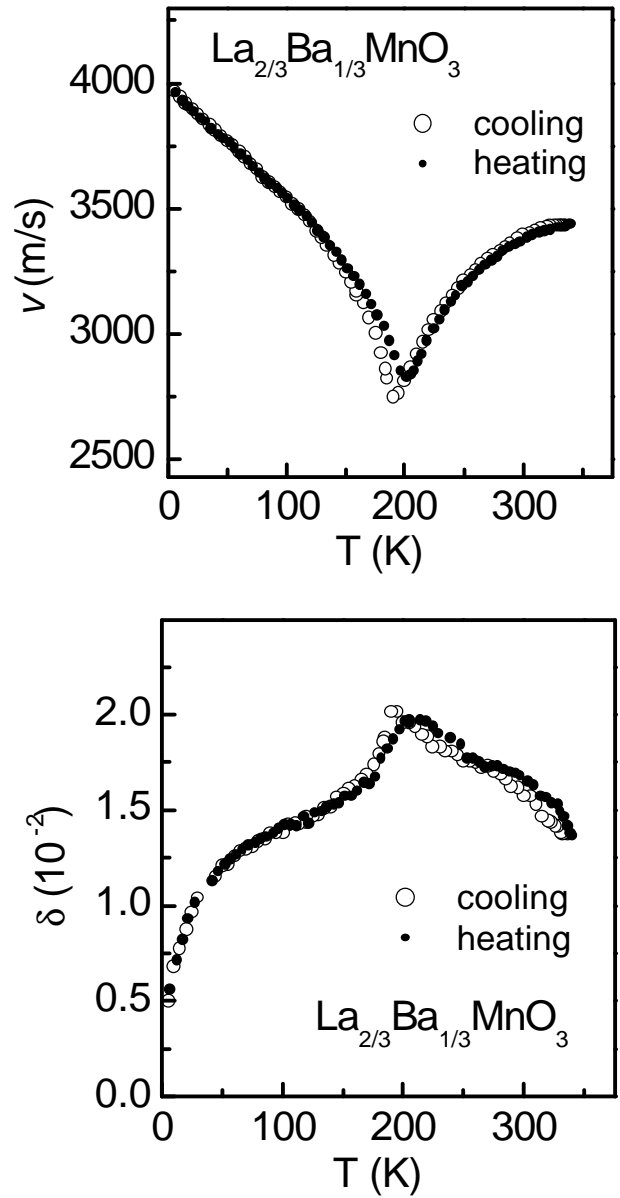

Fig. 1. Temperature dependences of the sound velocity *v* (upper panel) and logarithmic decrement δ (lower panel) of the $La_{2/3}Ba_{1/3}MnO_3$ manganite when cooling (o) and heating (•). Error bars are of the experimental points sizes approximately.

The lowest value of sound velocity and the corresponding highest value of sound damping correspond to equal coexisting phase fraction, i.e. to the region of the maximal rate of the interchange of the crystal phases.

The hysteresis loops of $v(T)$ and $\delta(T)$ (Fig.1) are asymmetrical: their extents to higher temperature (about 150 K) are much greater than down to lower temperature (about 55 K). They do not "converge" at 340 K, so the low temperature residual phase exists at this temperature. Thus the reverse martensitic transformation completes at higher temperature (about 370 K, as it can be supposed from Fig. 3, lower panel; see Sect. IV).

Acute minimum of $v(T)$ is similar to the temperature anomaly of the uniaxial pressure effect on the magnetic susceptibility $\chi(T)$, considered from phenomenological point of view as manifestation of an electronic transition.[14, 31]

As a whole the character of the temperature dependences obtained is ordinary for the martensitic compounds and alloys.[29,32] For example, ultrasound velocity vs temperature dependence in the ferromagnetic state of $Ni_{2.18}Mn_{0.82}Ga$[32] in the temperature region of the direct and reverse martensitic transitions is similar to one presented in Fig. 1 (upper panel). In this connection, it should be noted, that the dependences found have many common also with ones of $La_{1-x}Sr_xMnO_3$ (x=0.15-0.20) manganites, which experience a structural phase transition.[26,28] This similarity imply that the transition mentioned in $La_{1-x}Sr_xMnO_3$ has a martensitic nature as well, and so martensitic transition in $La_{2/3}Ba_{1/3}MnO_3$ is not something unique in the wide band manganites family.

## IV. CORRELATION OF THE STRUCTURE CHARACTERISTICS

### A. The phase separation process

As shown by neutron diffraction,[14, 23] $La_{2/3}Ba_{1/3}MnO_3$ experiences a structural transition from the rhombohedral $R\bar{3}c$ to the orthorhombic *Imma* space symmetry crystal phase. The difference between these crystal structures is seen in Fig. 2, where the positions of the Mn and O ions in the two lattices are presented. The transition between such symmetries has to be of the first order type (this follows from the group

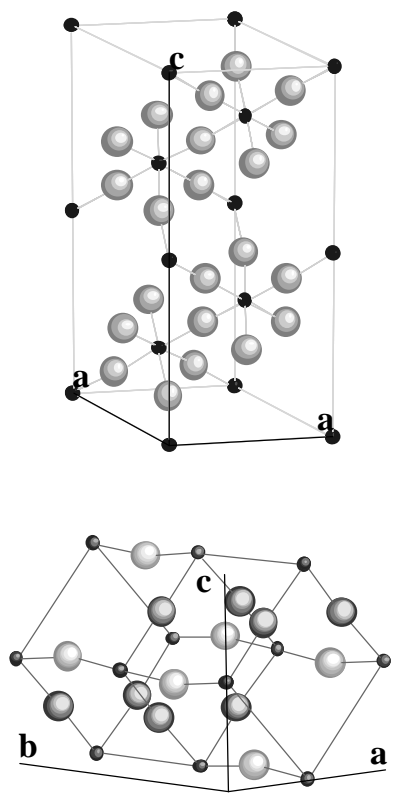

Fig. 2. Positions of the Mn (small circles) and O (large circles) ions in the rhombohedral $R\bar{3}c$ (upper panel) and orthorhombic *Imma* (lower panel) crystal structures of the La$_{2/3}$Ba$_{1/3}$MnO$_3$ compound. In the orthorhombic structure (lower panel) the large light circles represent the oxygen ions in the O1 positions (6 of 8 ions are shown), while the large dark circles represent ones in the O2 positions (8 of 12 ions are shown). La and Ba ions are not shown.

theory [33]), which is in accordance with our present ultrasound measurements (see Sec. III).

At 370 K only the rhombohedral $R\bar{3}c$ phase was observed. With decreasing temperature the orthorhombic *Imma* phase appeares (about 12% of *Imma* phase at 300 K) and gradually replaces the high temperature rhombohedral phase. But the high temperature $R\bar{3}c$ phase does not get replaced completely: its amount stays practically constant (of about 12%) below 140 K. The transition proceedes most rapidly at $T_S=200$ K, where the probability P(T)[*] of the simultaneous presence of the two phases in the $La_{2/3}Ba_{1/3}MnO_3$ compound is maximal (Fig. 3, lower panel). Since at this temperature the concentrations of both phases are equal and the square of the surface separating the phases has its maximal value, inner stresses caused by the boundaries between phases are maximal as well. The last must lead to an increase of the diffraction lines halfwidths. This is confirmed by the experiment: as shown in Fig. 3 (upper panel) the line halfwidth for the diffraction angle $2\theta=126°$ (at this angle the resolution of the diffractometer is best) increases sharply in both phases when approaching 200 K.

The large extension of the region of the phase interchange is characteristic for a martensitic phase transformation: the $R\bar{3}c$ and *Imma* phases were found to coexist in the whole region below 370 K.

### B. Temperature anomalies of structure parameters

Additional contribution to the temperature dependence of the inner stresses in the compound, discussed in the Sec. A, is caused by a difference in the thermal expansion coefficients of the coexisting phases. As seen in Fig. 4 (upper panel) the volume thermal expansion coefficient of the high-temperature rhombohedral phase increases abnormally in the vicinity of $T_s=200$ K, caused by the first order phase transition in this phase.[14] The main contribution to this anomaly is brought by the thermal expansion of the *c* parameter of the $R\bar{3}c$ crystal lattice (Fig. 4, lower panel). The low-temperature orthorhombic phase has no transition at 200 K, as one can see from Fig. 4 (upper panel), but nevertheless it presents an anomaly at ~240 K. Moreover, the *Imma* phase exhibits clear anomalies in the temperature dependence of internal 11structure parameters: as seen in Fig. 5 (upper panel) the temperature dependence of the angles between the Mn and O ions in different positions (see Fig. 2) in the crystal lattice exhibits singularities at 230 K (sharp bends) and 250 K (break). Note, that an anomaly of the sharp bend type is observed in the $R\bar{3}c$ phase in the $T_s$ region as well: the $c/a$ ratio of the crystal lattice parameters has a break of the derivative at 200 K (Fig. 5, lower panel), and the Mn-O-Mn bond angle has it at 190 K (Fig. 5, upper panel).

---

[*] The ordinate axis in the lower panel of Fig. 3 presents the normalized probability P of the simultaneous presence of the two phases in the $La_{2/3}Ba_{1/3}MnO_3$ compound. The value is defined as $P=P_1P_2/(P_1P_2)_{max}$, where $P_1$ and $P_2$ are the atomic concentrations of the phases in the system.

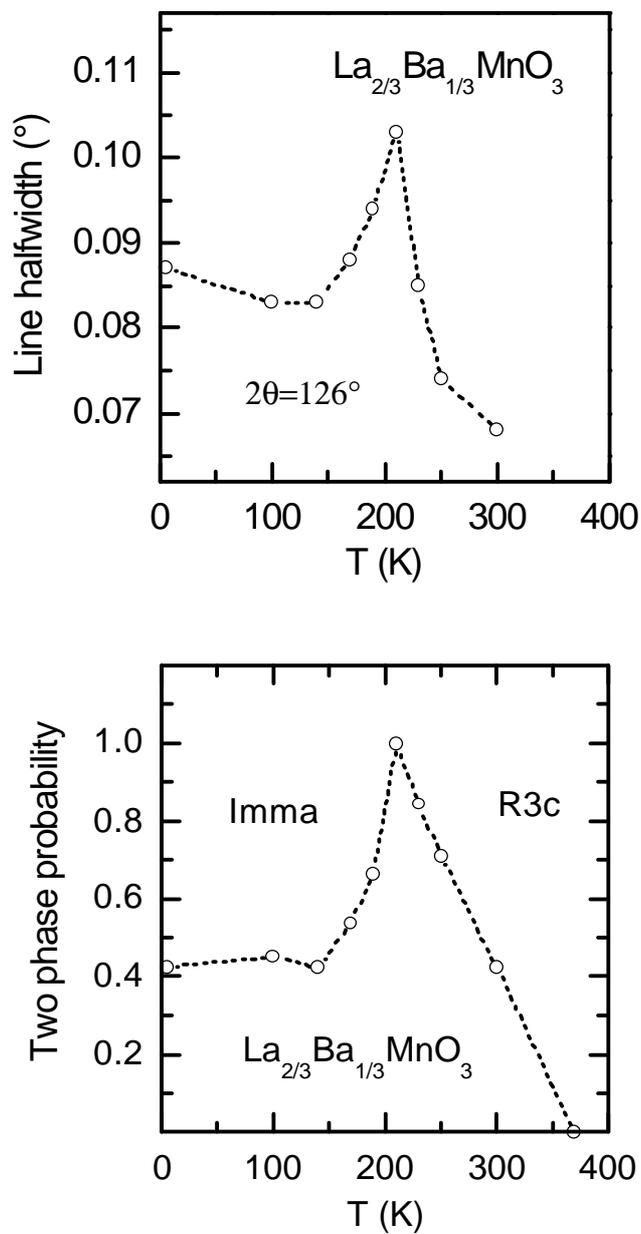

Fig. 3. Temperature dependences of the line halfwidth for the diffraction angle $2\theta=126°$ for the $R\bar{3}c$ and *Imma* phases (upper panel), and of the normalized probability of the simultaneous presence of the two phases in the $La_{2/3}Ba_{1/3}MnO_3$ compound (lower panel). Dotted lines are guides to the eyes, error bars are of the experimental points sizes approximately.

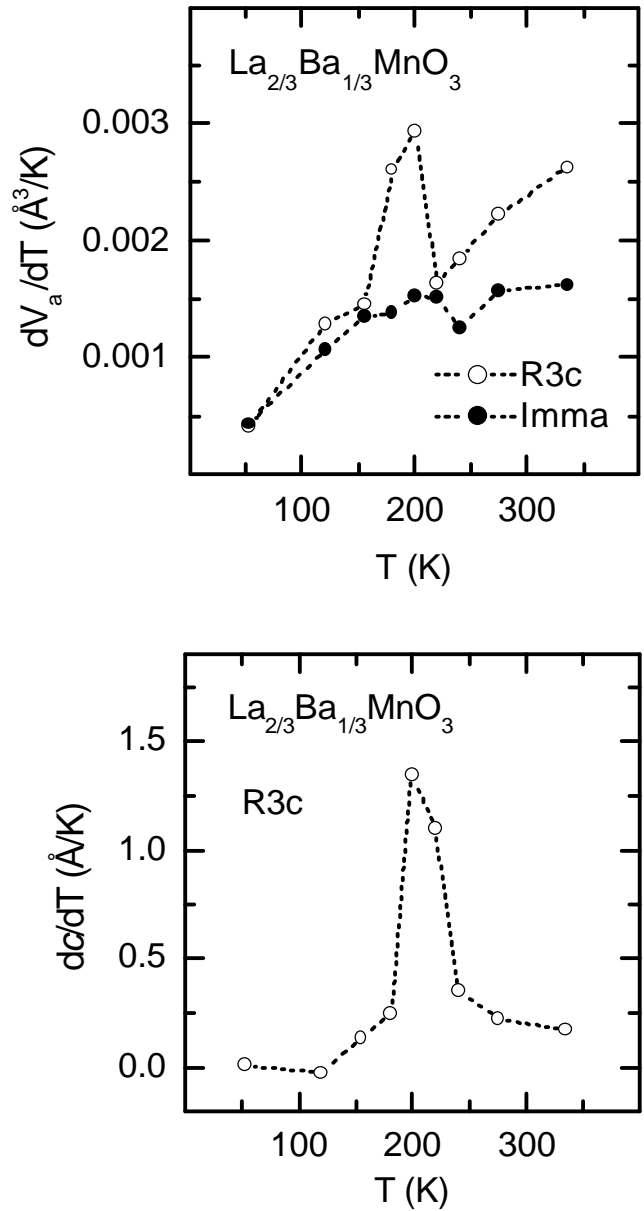

Fig. 4. Temperature derivatives of the volume $V_a$ per manganese atom for the $R\bar{3}c$ and *Imma* phases (upper panel), and of the $c$ parameter (doubled Mn-Mn distance $d$) of the $R\bar{3}c$ crystalline phase (lower panel) in the $La_{2/3}Ba_{1/3}MnO_3$ compound. Dotted lines are guides to the eyes, error bars are of the experimental points sizes approximately.

    Let us now consider the oxygen subsystem in more detail. The crystal structures of $La_{1-x}Ba_xMnO_3$ manganites can be described as distorted perovskite $ABO_3$ type structures (A=$La_{1-x}Ba_x$). The A cations occupy cavities of the three-dimensional network of vortex sharing $BO_6$ octahedra. The size of the cavity in the undistorted perovskite structure depends on the B-O bond length. A first type of distortion consists in a tilting of the $BO_6$ octahedra, which modifies the size of the cavity available for the A-type cation. Different systems of tilting correspond to different superstructures of the cubic perovskite type.[34, 35, 36] A second type of distortion observed in perovskite manganites is the Jahn-Teller (JT) distortion of the $MnO_6$ octahedral groups, which is characterized by two long and four short Mn-O bonds.[35, 36] The symmetry of the rhombohedral structure does not allow such a distortion, but the orthorhombic one is favorable for it. As one can see in Fig. 6 such a distortion exists in the *Imma* phase and has a nontrivial temperature dependence. The Mn-O1 bond length has a sharp bend at 190 K (as well as the Mn-O bond length in the $R\bar{3}c$ phase) and a break between 230 K and 250 K, while the Mn-O2 bond length has a weakly expressed bend at 230 K. Note, that the behavior of the Mn-O-Mn bond angles between 190 K and 5 K substantially differs from the behavior of the bond lengths, as clearly seen in Fig. 6.

    Thus, the *Imma* crystal structure contains small coherent JT distortions. The bond lengths found and the small value of the JT distortion are in a good agreement with earlier results.[34] The temperature behavior of the bond distances and angles (Fig. 4- Fig. 6) implies, that the low temperature *Imma* phase experiences an electronic transition above 200 K (between 210 and 250 K).[37]

### C. Electronic transition.

    The found correlation of the various characteristics related to the system as a whole and to both observed phases appears to be rather unexpected. To analyze the situation let us consider Gibbs free energies[38] $\Phi_1(x_1,…x_i,…x_n,p,T)$ and

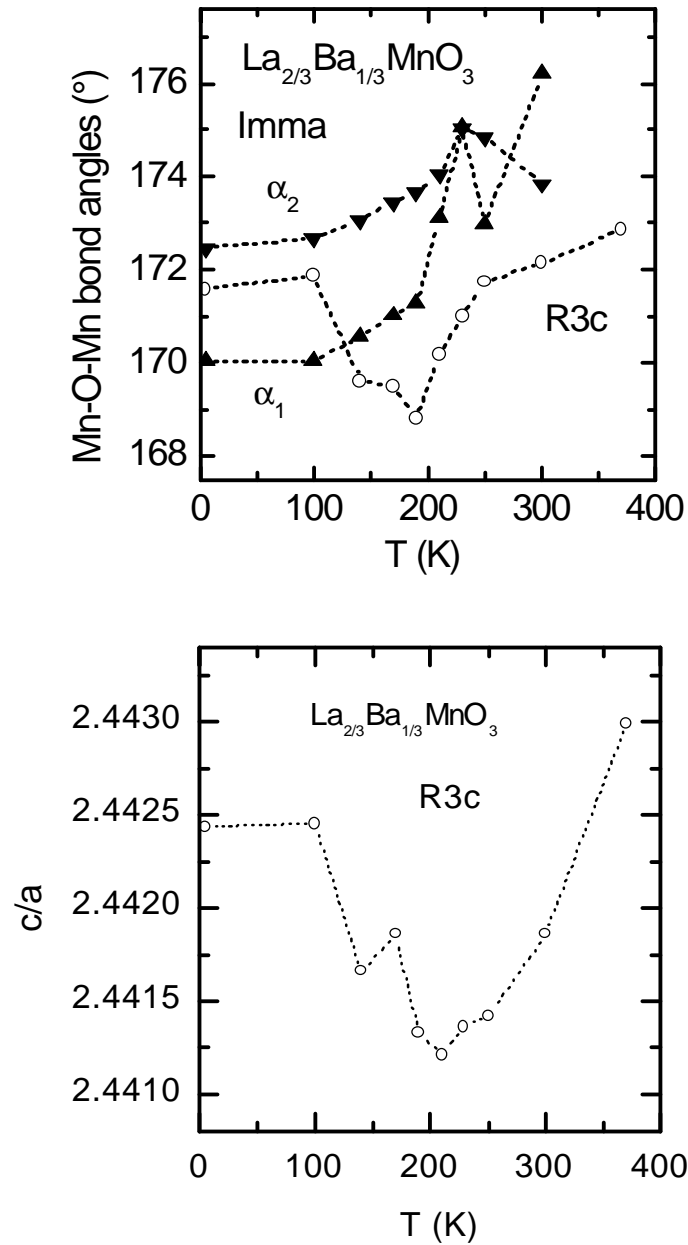

Fig. 5. Mn-O-Mn bond angles as a function of temperature for rhombohedral $R\bar{3}c$ (circles) and orthorhombic *Imma* (▲-$\alpha_1$, ▼-$\alpha_2$) phases of $La_{2/3}Ba_{1/3}MnO_3$ (upper panel, $\alpha_{1(2)}$ stands for the Mn-O1(2)-Mn angle); temperature dependence of the $c/a$ ratio of the $R\bar{3}c$ phase lattice parameters (lower panel). Dotted lines are guides to the eyes, error bars are of the experimental points sizes approximately.

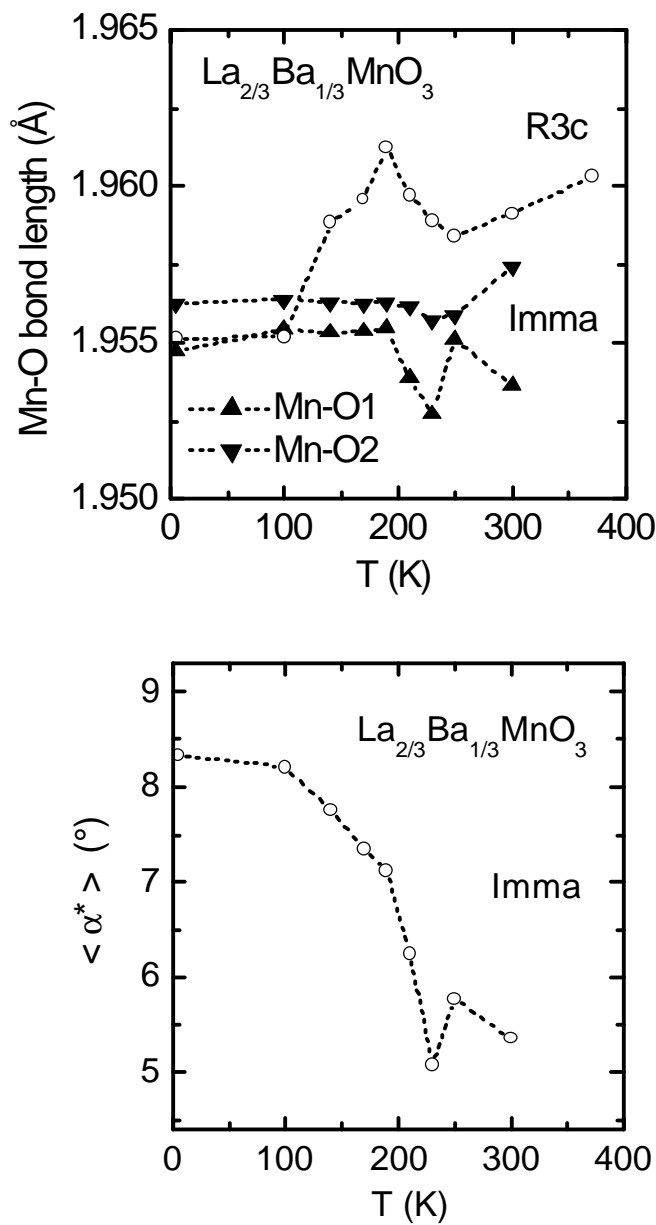

Fig. 6. Mn-O bond lengths as a function of temperature for rhombohedral $R\bar{3}c$ (circles) and orthorhombic *Imma* (▲- Mn-O1, ▼- Mn-O2) phases of $La_{2/3}Ba_{1/3}MnO_3$ (upper panel); average deviation from 180° of the Mn-O-Mn bond angle for the *Imma* phase $<\alpha^*>=180°-(\alpha_1+2\alpha_2)/3$ (lower panel). Dotted lines are guides to the eyes, error bars are of the experimental points sizes approximately.

$\Phi_2(y_1,\ldots y_j,\ldots y_m,p,T)$ of the high and low temperature quasiequilibrium phases, respectively (here $x_i$ and $y_j$ are parameters of the system, the total number of which is n or m, $p$ is an external parameter, and $T$ is temperature). Minima of $\Phi_1$ and $\Phi_2$ on $x_i$ and $y_j$ determine trajectories $x_i(p,T)$ and $y_j(p,T)$. Conditions

$$\Phi_1[x_1(T),\ldots x_i(T),\ldots x_n(T),p,T] < \Phi_2[y_1(T),\ldots y_j(T),\ldots y_m(T),p,T] \text{ at } T > T_s,$$

$$\Phi_1[x_1(T),\ldots x_i(T),\ldots x_n(T),p,T] > \Phi_2[y_1(T),\ldots y_j(T),\ldots y_m(T),p,T] \text{ at } T < T_s \quad (1)$$

determine a temperature phase transition

$$\Phi_1[x_1(T),\ldots x_i(T),\ldots x_n(T),p,T] \to \Phi_2[y_1(T),\ldots y_j(T),\ldots y_m(T),p,T]$$

at the temperature $T_s$, so discontinuities of the derivatives of the free energy of the system occurs at this temperature (Fig. 7). But conditions (1) do not require the temperature derivatives of $\Phi_1$ and $\Phi_2$ or their parameters to have any singularities in a vicinity of $T_s$.

In martensitic transformations both phases (the high temperature high symmetry and the low temperature low symmetry ones) are present simultaneously in the system. Generally speaking they are in the long-living metastable states, but for a number of processes (which proceed not too slow) they may be considered as being in thermally equilibrium states. So, let us consider next situation:

$$\Phi_1[x_1(T),\ldots x_i(T)\ldots x_n(T),p,T] < \Phi_2[y_1(T),\ldots y_j(T)\ldots y_m(T),p,T] \text{ at } T > T_s,$$

$$\Phi_1[x_1(T),\ldots x_i(T)\ldots x_n(T),p,T] > \Phi_2[y_1(T),\ldots y_j(T)\ldots y_m(T),p,T] \text{ at } T < T_s,$$

$$\Phi_1'[x_1'(T),\ldots x_i'(T)\ldots x_n'(T),p,T] > \Phi_2[y_1(T),\ldots y_j(T)\ldots y_m(T),p,T] \text{ at } T > T_s, \quad (2)$$

$$\Phi_1'[x_1'(T),\ldots x_i'(T)\ldots x_n'(T),p,T] > \Phi_1[x_1(T),\ldots x_i(T)\ldots x_n(T),p,T] \text{ at } T > T_s',$$

$$\Phi_1'[x_1'(T),\ldots x_i'(T)\ldots x_n'(T),p,T] < \Phi_1[x_1(T),\ldots x_i(T)\ldots x_n(T),p,T] \text{ at } T < T_s',$$

where $\Phi_1'$ is an additional metastable phase, whereas $\Phi_1$ is the thermally equilibrium high symmetry phase. Hence when lowering temperature we have phase transition between the high symmetry and low symmetry phases

$$\Phi_1[x_1(T),\ldots x_i(T)\ldots x_n(T),p,T] \to \Phi_2[y_1(T),\ldots y_j(T)\ldots y_m(T),p,T]$$

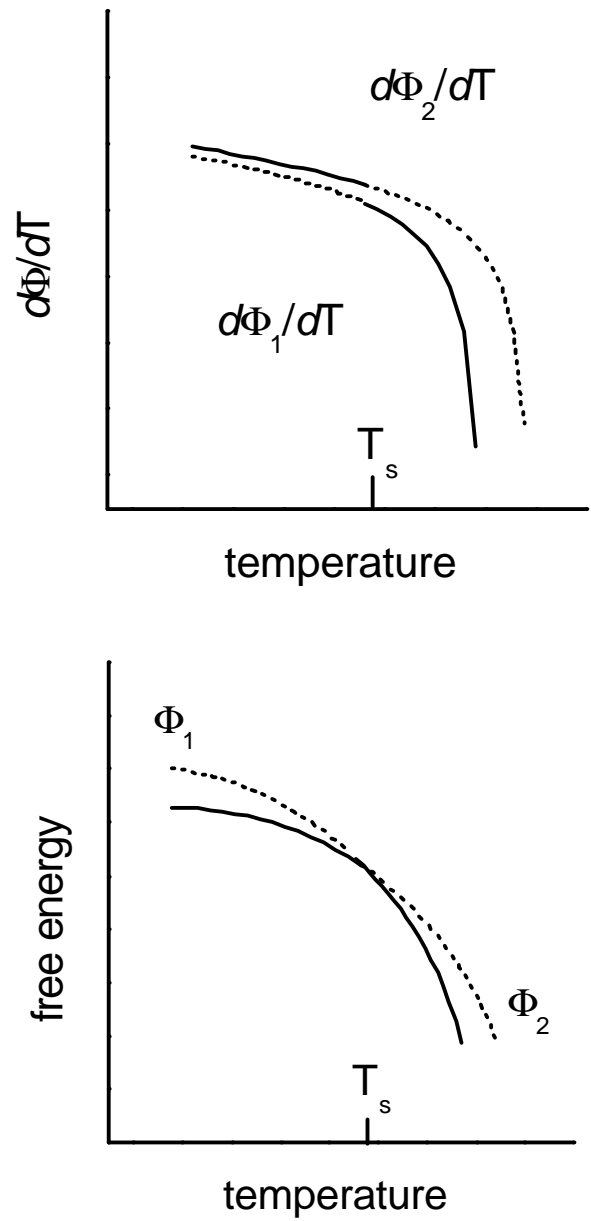

Fig. 7. Temperature dependences of the Gibbs free energies of the high temperature ($\Phi_1$) and low temperature ($\Phi_2$) phases (lower panel) and their temperature derivatives (upper panel) at the first order phase transition. Solid line corresponds to the thermodynamically equilibrium state, $T_s$ indicates the transition temperature.

at $T = T_s$, and a transition within the high symmetry phase (metastable at $T < T_s$)

$$\Phi_1[x_1(T),...x_i(T)...x_n(T),T] \to \Phi'_1[x'_1(T),...x'_j(T)...x'_m(T),T]$$

at $T = T'_s$, where the $T'_s$ temperature is some lower than $T_s$. Thus we have two phase transitions at the temperatures close one to another, but since conditions (2) do not require the temperature derivatives of $\Phi_2$ or its parameters to have any singularities around $T_s$, the question about the correlation of characteristics of the system and the separating phases remains.

If we take, however, into account the interaction between the phases, which is especially strong around the $T_s$ temperature, as it was mentioned in the Sections A and B, the free energies of the "perturbed" phases can be presented as

$$\Phi_1^*[x_1(T),...x_i(T)...x_n(T), y_1(T),...y_j(T)...y_m(T), p, T] = \Phi_1[x_1(T),...x_i(T)...x_n(T), p, T]$$

$$+ \delta\Phi_1[x_1(T),...x_i(T)...x_n(T), y_1(T),...y_j(T)...y_m(T), p, T],$$

(3)

$$\Phi_2^*[y_1(T),...y_j(T)...y_m(T), x_1(T),...x_i(T)...x_n(T), p, T] = \Phi_2[y_1(T),...y_j(T)...y_m(T), p, T]$$

$$+ \delta\Phi_2[y_1(T),...y_j(T)...y_m(T), x_1(T),...x_i(T)...x_n(T), p, T],$$

where the additional terms $\delta\Phi_1$ and $\delta\Phi_2$ are caused by the interaction, and $x_i(T)$ and $y_j(T)$ are new selfconsistent trajectories. The additional terms $\delta\Phi_1$ and $\delta\Phi_2$ increase quickly in the vicinity of the transition temperature $T_s$, so we could get correlated anomalies of the characteristics of the system and both phases. Such anomalies (generated by the interaction between the phases) should reflect the temperature dependence of the interaction between the phases, which is not the case.[†]

---

[†] Phase transitions in the coexisting phases, caused by their interaction, should occur right-hand $T_s$, (in cooling process) when the interaction exceeds certain threshold value (corresponding to certain level of the line halfwidth in Fig. 3), and then left-hand $T_s$ a reentrant transition have to occur, when the interaction drops below this threshold value. However, as it follows from Fig. 6 (lower panel) and Fig. 9 (lower panel), as well as from Fig. 12 in Ref.[14], there are no reentrant transitions in the both coexisting phases.

Thus, taking into account the negative conclusions of the analysis made above, we suppose, that the reason for the correlated transitions between the phases $R\bar{3}c$ and *Imma* (Fig. 3) and within the $R\bar{3}c$ phase (Fig. 4, Fig. 5), as well as of the anomalies in the *Imma* phase (Fig. 5, Fig. 6) can be interconfiguration electronic transition within the "molecule" (i.e. electrically neutral group of the ions, forming the formula unit of the compound) $La_{2/3}Ba_{1/3}MnO_3$. On such a scenario an electron transition in the "molecule" occurs when its volume reaches certain value in the cooling process (note, that the "molecular" volumes of the $R\bar{3}c$ and *Imma* phases are equal at $T_s$).[14] The crystal lattice reflecting this transition proceeds from the $R\bar{3}c$ to the more energy favorable Jahn-Teller *Imma* phase. In the coexisting metastable $R\bar{3}c$ and *Imma* lattices the electronic transition in the "molecule" $La_{2/3}Ba_{1/3}MnO_3$ (i.e. a sharp change of the spatial electron density distribution, which minimizes the energy at new values of external parameters of the system) leads to the jump of the formula unit volume for the $R\bar{3}c$ phase[14] and to the sharp decrease of the Mn-O-Mn bond angles for the *Imma* phase. So, the "molecular" (i.e. independent on crystal structure) electronic transition appears to be the driving motive of the structural transformation considered.

## V. SPIN-LATTICE COUPLING

The character of the electronic states determines the lattice as well as the magnetic behavior of the system. The spin-lattice coupling (some features of the latter in manganites are considered in Ref. [39]) will develop most strongly in the magnetically ordered state (at T 340 K, depending on the oxygen index $\delta$ in $La_{2/3}Ba_{1/3}MnO_{3-\delta}$ [40]). The magnetic ordering occurs in the both phases simultaneously,[14] but is reflected in different manner in their crystal parameters. Although the $R\bar{3}c$ phase is the main phase in the $T_C$ region its lattice parameters react less markedly on the appearance of the magnetic order.[14] The *a* and *b* parameters of the *Imma* phase react also weakly on the magnetic order parameter, whereas the *c* parameter experiences a strong spontaneous magnetostriction.[14] As one can see from Fig. 8 (lower panel) a strong correlation exists between the increasing lattice distortion (as estimated by the difference *c-a*) and the intensity of the magnetic neutron scattering peak,[14] which is proportional to the square of the magnetic order parameter.[41] On the other side, the magnetic moment of the Mn ions correlates as well with changes of the interatomic distances at the first order phase transition in the $R\bar{3}c$ phase at $T_s$, as it can be seen in Fig. 8 (upper panel), where the experimentally determined dependence of the Mn-Mn distance *d* in this phase and a dependence corresponding to the Debye-Grüneisen model (see Appendix A) are presented. The non Debye-Grüneisen contribution $\Delta d$ to the Mn-Mn distance as a function of temperature is depicted in Fig. 9 (lower panel), and the corresponding contribution $\Delta I^*_{magn}$ to the integrated intensity *I* of magnetic Bragg peak (see Appendix B) is presented in the upper panel of Fig. 9.‡ A clear correlation between the $\Delta d$ and $\Delta I^*_{magn}$ is evident from these data.

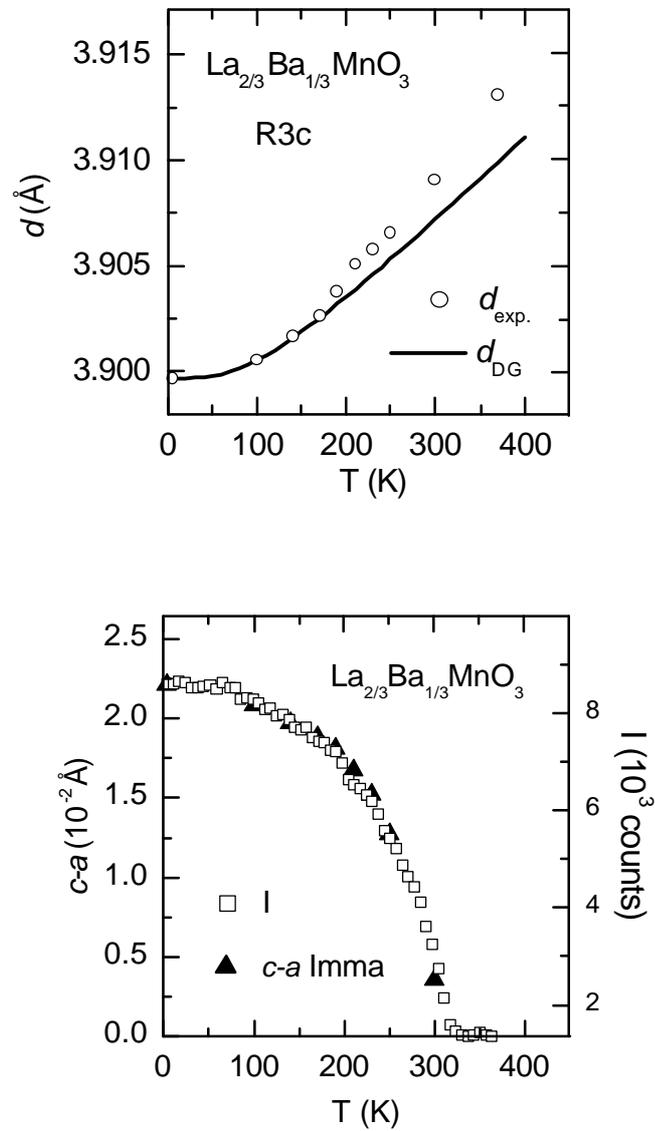

Fig. 8. Temperature dependences of the Mn-Mn distance $d$ for the $R\bar{3}c$ crystalline phase of the $La_{2/3}Ba_{1/3}MnO_3$ compound (upper panel, circles depict the experimental data and the solid line depicts the Debye-Grüneisen curve), of the $c$-$a$ difference for the *Imma* phase and of the integrated intensity $I$ of magnetic Bragg peak (triangles and squares, respectively; lower panel). Error bars are of the experimental points sizes approximately.

## VI. Microstructure effects

As it was mentioned above martensitic transitions are first order, solid-solid, diffusionless structural phase transformations. Unlike other phase transformations where diffusion disperses the neighboring atoms, here, neighboring atoms in the parent phase also remain in the product phase.[42] However, the lattice gets distorted due to spontaneous displacement of the atoms (from their positions in the parent lattice) accompanying the discontinuous change in the shape and symmetry of the unit cell. This creates long-range strain fields, which in turn strongly depend on the relative positions and orientations of the martensitic plates. The transformation path depends on continuously evolving configuration-dependent long-range strain fields that eventually leave the system in a metastable state. Typically, the martensite morphology consists of thin plate-like domains with twinned structure that are oriented along elastically favorable habit plane directions. Pictures of this sort may be observed by the electron microscopy technique. We have performed a corresponding study of the microstructure characteristics of $La_{2/3}Ba_{1/3}MnO_3$.

TEM results were obtained on $La_{2/3}Ba_{1/3}MnO_3$ both at room temperature (300 K) and at low temperature (100 K). The crystal flakes investigated by TEM are micron size large and the transparent thickness of the sample for 300 or 400 kV electrons is around 200-300 nm. For HREM, the maximum thickness is around 50 nm.

The electron diffraction study did not show any clear evidence of a microstructural difference between the $R\bar{3}c$ room temperature phase and the *Imma* low temperature phase. The small difference between the lattice parameters of both phases[14] and the fact that electron diffraction (ED) is very sensitive to multiple diffraction, makes that the ED patterns look very similar at 100 K and 300 K. Three ED patterns along $[\bar{1}11]^*$, $[\bar{2}01]^*$, and $[14\ 7\ \bar{9}]^*$ obtained at room temperature are shown in Fig. 10. These ED patterns also match perfectly with the [111]*, [121]* and [032]* ED patterns of the *Imma* low temperature phase, respectively. Actually only small differences in intensity should be indicative of a phase transformation.

---

‡ Value $\Delta I^*_{magn}$ characterizes a difference between the magnetic system parameters left-hand and right-hand from $T_s$.

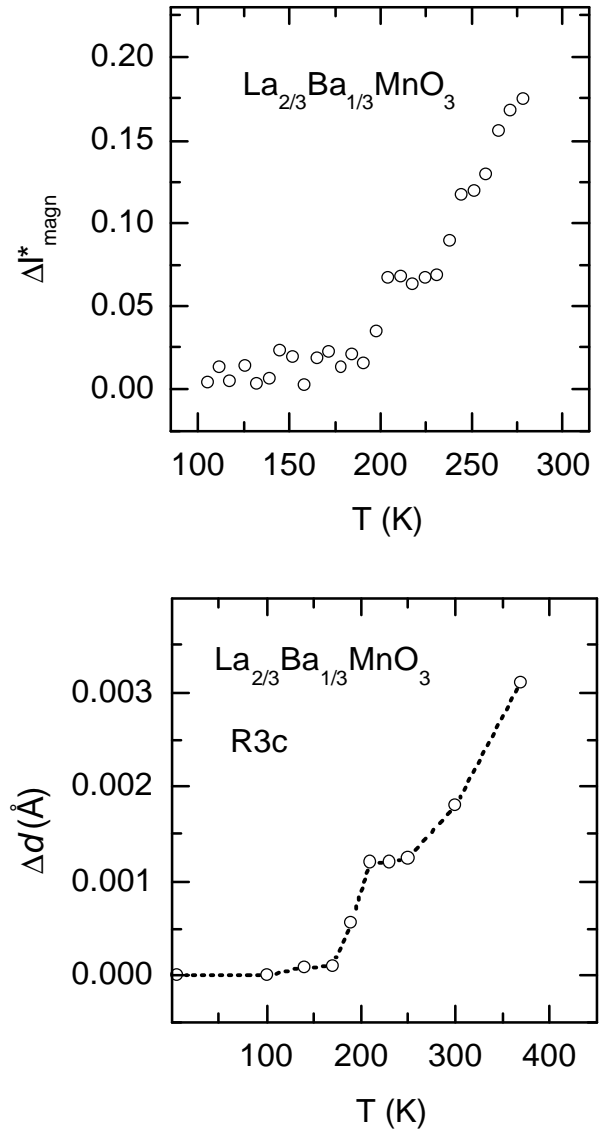

Fig. 9. The first order phase transition contributions $\Delta I^*_{magn}$ (to the integrated intensity $I$ of magnetic Bragg peak, upper panel), and $\Delta d$ (to the Mn-Mn distance $d$ in the $R\bar{3}c$ crystalline phase, lower panel) of the $La_{2/3}Ba_{1/3}MnO_3$ compound. Dotted line is guide to the eyes, error bars are of the doubled experimental points sizes approximately.

In real space, a tweed structure indicative of a first-order (martensitic) phase could be expected upon cooling, but was never observed. At room temperature most grains are single crystalline with very little defects. Occasionally however the presence of planar defects running along {012}- and $\{1\ \bar{1}2\}$-type planes is observed (Fig. 11). These defects, indicated by arrowheads, can be identified as antiphase boundaries and do not disappear or change upon cooling the sample to 100K.

The electron microscopy study of $La_{2/3}Ba_{1/3}MnO_3$ samples was unable to detect any structural phase transformation between 100 and 300 K. A possible reason of this peculiar behavior is the film geometry that leads to a difference in the relaxation processes and a shift of the martensitic transition temperature. The $La_{2/3}Ba_{1/3}MnO_3$ compound as well as other compounds exhibiting a martensitic phase transformation are extremely sensitive to sample preparation history and measurement protocol.[9,11,43] I.e., the martensitic phase transition is an intrinsic feature of a large number of mixed valence manganites, but the observation of the effects connected with the transition depends on a number of factors. Sample size is one among them. Recently it was shown, that ultrasharp magnetization steps in phase-separated manganites, having the martensitic nature, appear only in thick (>0.5 μm) films.[11] Besides, the size dependence of structure and magnetic properties of $La_{2/3}Sr_{1/3}MnO_3$ nanoparticles was studied.[44] It was established that as the grain size *D* decreases, the crystal symmetry changes from rhombohedral to cubic at *D*~16 nm. Our HREM sample thickness in the thinnest part is indeed in this range but ED patterns are taken over much thicker regions and moreover the crystals are only small in one dimension, compared to the measurements in.[44]

Another possible reason is related to the fact that the samples for electron microscopy were prepared from the bulk sample studied by neutron diffraction technique earlier, so they experienced several full thermal cycles. But most grains show very few defects. The last implies the thermoelastic character of the martensitic transformation in the compound studied, as it is typical when a difference between martensite and austenite lattice parameters is small.[45] A final possible reason is related to the electron irradiation of the sample. 400 kV irradiation may be responsible for a lowering of the martensitic transition temperature.

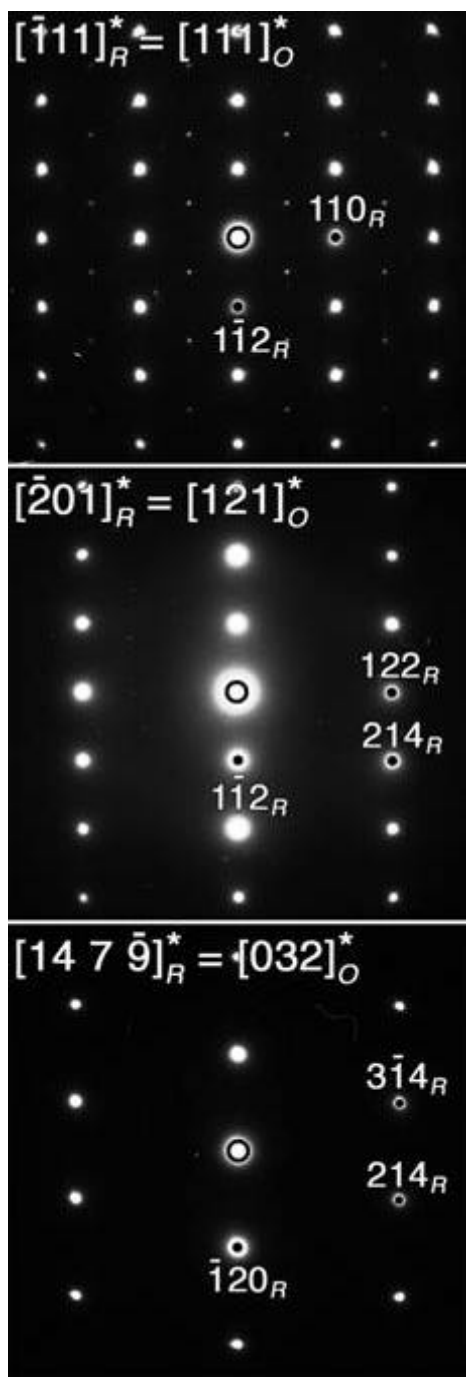

Fig. 10. ED patterns of $La_{2/3}Ba_{1/3}MnO_3$ along $[\bar{1}11]_R^*$, $[\bar{2}01]_R^*$, and $[14\ 7\ \bar{9}]_R^*$ obtained at 300 K (*R* – rhombohedric phase). These ED patterns also match perfectly with the $[111]_O^*$, $[121]_O^*$ and $[032]_O^*$ ED patterns of the *Imma* low temperature phase (*O* – orthorhombic phase), respectively.

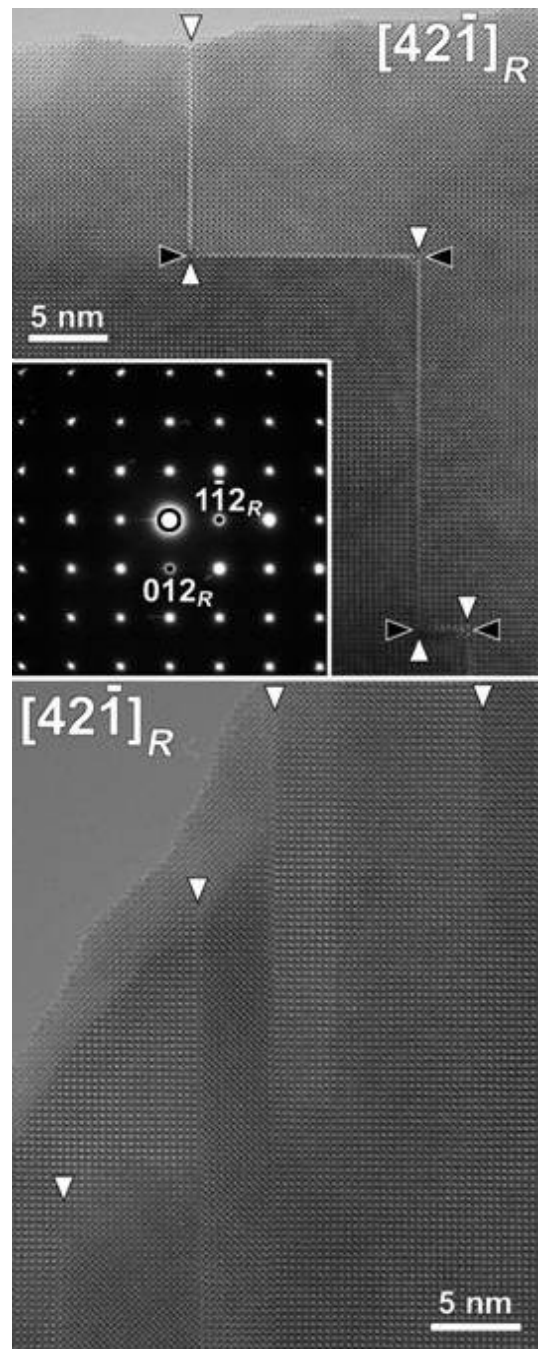

Fig. 11. Room temperature HREM images along the $[42\bar{1}]$ zone. Note the presence of planar defects running along the {012}- and {1$\bar{1}$2}-type planes (indicated by arrowheads). The corresponding ED pattern of the top HREM image is shown as an inset.

## VII. SUMMARY

Structural phase transformation in the polycrystalline $La_{2/3}Ba_{1/3}MnO_3$ compound has been studied by ultrasound, neutron diffraction and electron microscopy techniques and the following conclusions have been made:

1) Minima of the velocity $v$ and maxima of the decrement $\delta$ of the ultrasound wave in the cooling and heating processes detect a first order phase transition with the characteristic temperature $T_s$=200 K.

2) Extended temperature hysteresis and giant alteration of $v$ and $\delta$ prove martensitic character of the phase transition at macroscopic level.

3) Electronic structure of the coexisting $R\bar{3}c$ and *Imma* phases of the compound exhibits anomalous behavior in the vicinity of $T_s$: temperature dependences of the Mn-O-Mn bond angles and Mn-O bond lengths have singularities in the temperature region 190-230 K.

4) The $R\bar{3}c$ and *Imma* phases parameters have substantially different character of the correlation with parameters of the magnetic subsystem of the compound.

5) Low level of the defect concentration in the system, experienced several full thermal cycles, and the small difference between the lattices parameters of the high and low temperature phases imply the thermoelastic character of the martensitic transformation in $La_{2/3}Ba_{1/3}MnO_3$.

6) The martensitic transition in $La_{2/3}Ba_{1/3}MnO_3$ is tightly connected with electronic transitions in the both coexisting crystal phases.

**Acknowledgements**

Authors are grateful to G. Van Tendeloo, A.I. Prokhvatilov, V.I. Fomin and N.N. Galtsov for interest in the work and fruitful discussions.

## APPENDIX A

The $\Delta d$ value is determined as $\Delta d = d_{exp} - d_{DG}$, where $d_{exp}$ is the experimental value, and the Debye-Grüneisen term $d_{DG}$ has form[38]

$$d_{DG} = d_0 + p\left(\frac{T}{q_D}\right)^4 \cdot \int_0^{\frac{q_D}{T}} \frac{x^3 \cdot dx}{\exp(x)-1}. \qquad (A1)$$

Debye temperature $\theta_D$ was estimated from the present experimental data as[46]

$$\mathbf{q}_D = \frac{\hbar}{k_B} \cdot q_D \cdot v_s. \qquad (A2)$$

The formula unit volume $V_a = 59.35$ Å$^3$ at T=5 K was used to calculate the Debye wave number in Eq. (A2)

$$q_D = \left(\frac{6p^2}{V_a}\right)^{1/3}.$$

The sound velocity in the infinite substance $v_s$ in Eq. (A2) was estimated as[47]

$$v_s = v\sqrt{\frac{1-\mathbf{s}}{(1-\mathbf{s})(1-2\mathbf{s})}},$$

with the sound velocity in the rod $v = 3.97 \cdot 10^3$ m/s at T=5 K, and roughly estimated Poisson coefficient $\sigma \approx 0.33$. Then the two fitting parameters $d_0$ and $p$ in Eq. (A1) were determined using two low temperature experimental points (less perturbed by the phase transition) 5 and 100 K (Fig. 8, upper panel).

## APPENDIX B

The $\Delta I^*_{mag}$ value is determined as $\Delta I^*_{mag} = I^*_{mag} - I_{fit}$. The $I^*_{mag}$ value is derived from the experimental data as

$$I^*_{mag} = \frac{I - I_{para}}{I_0 - I_{para}},$$

where $I$ is the experimental values of the magnetic Bragg peak (Fig. 8, lower panel), $I_{para}$ is the peak intensity in the paramagnetic region, and $I_0$ is the maximal value of the magnetic peak intensity. The $I_{fit}$ value is defined as a root of the equation

$$I_{fit} = B_S\left(\frac{3S}{S+1} \cdot \frac{T^*_C}{T} \cdot I_{fit}\right), \qquad (B1)$$

where $B_S$ is the Brillouin function[48] with S=3/2 and $T_C^*$=350 K. This value is close to the experimental data on the magnetic neutron scattering below $T_s$. Note, that here we chose the $I_{fit}$ curve in the (B1) form just for a convenience without any discussion on its physical content.

*Corresponding author. Elena L. Fertman, Institute for Low Temperature Physics and Engineering, 47 Lenin Avenue, Kharkov 61103, Ukraine.

Email address: Fertman@ilt.kharkov.ua; FAX: 38-057-3450593.